\documentclass[12pt]{article}

\usepackage{amsmath}
\usepackage{feynmf}
\usepackage{latexsym}
\usepackage{amssymb}
\usepackage[dvips]{graphicx}
\usepackage{caption}
\usepackage{rotating}
\usepackage{latexsym}
\usepackage{verbatim}

 \def\be{\begin{equation}}
 \def\ee{\end{equation}}
 \def\bea{\begin{eqnarray}}
 \def\eea{\end{eqnarray}}
 \def\bei{\begin{itemize}}
 \def\eei{\end{itemize}}
 \def\bs{\begin{slide}}
 \def\es{\end{slide}}
 \def\nn{\nonumber}
 
 \def\pd{\partial}
 
 \def\L{\mathcal{L}}
 \def\M{\mathcal{M}}
 \def\a{\alpha}

 \def\g{\gamma}

 \def\D{\Delta}
 \def\m{\mu}
 \def\n{\nu}
 
 \def\r{\rho}
 \def\th{\theta}

 \def\l{\lambda}

 \def\s{\sigma}
 
 \def\sb{\bar{\s}}
 \def\psib{{\bar{\psi}}}
 
 \def\f{\phi}

 \def\MZp{M_{Z'}}
 \def\MZO{M_{Z_0}}
 
 \def\({\left(}
 \def\){\right)}
 \def\[{\left[}
 \def\]{\right]}

 \def\la{\langle}
 \def\ra{\rangle}
 \def\Tr{\textnormal{Tr}}

 \def\thth{\theta^2 \bar\theta^2}

 \def\QHu{{Q_{H_u}}}
 \def\QHd{{Q_{H_d}}}
 \def\cA{\mathcal{A}}

 \def\numeq{n^\text{eq}}
 \newcommand{\bth}{{\bf 3}}
 \newcommand{\btw}{{\bf 2}}
 \newcommand{\bon}{{\bf 1}}

 \def\ds#1{#1\kern-1ex\hbox{/}}
 \def\sla{\raise.15ex\hbox{$/$}\kern-.57em}
 \def\Stuckelberg{St\"uckelberg }

\begin{document}

\begin{titlepage}

\rightline{ROM2F/2008/25}
%\rightline{DAMTP-2008-107}

\vskip 2cm

\centerline{{\large\bf St\"uckelino Dark Matter in Anomalous
$U(1)'$ Models}}

\vskip 1cm

\centerline{Francesco
Fucito\footnote{Francesco.Fucito@roma2.infn.it}$^\natural$,
            Andrea
Lionetto\footnote{Andrea.Lionetto@roma2.infn.it}$^\natural$,
            Andrea
Mammarella\footnote{Andrea.Mammarella@roma2.infn.it}$^\natural$,
            Antonio
Racioppi\footnote{Antonio.Racioppi@kbfi.ee}$^\flat$}

\vskip 1cm

\centerline{$^\natural$ Dipartimento di Fisica dell'Universit\`a
di Roma , ``Tor Vergata" and} \centerline{I.N.F.N.~ -~ Sezione di
Roma ~ ``Tor Vergata''}
%\vskip .2 cm
\centerline{Via della Ricerca  Scientifica, 1 - 00133 ~ Roma,~
ITALY}
\vskip 0.25cm
\centerline{$^\flat$ National Institute of Chemical Physics and
Biophysics,}
%\vskip .2 cm
\centerline{Ravala 10, Tallinn 10143, Estonia}
\vskip 1cm

\begin{abstract}
We study a possible dark matter candidate in the framework of a
minimal anomalous $U(1)'$ extension of the MSSM. It turns out that
in a suitable decoupling limit the St\"uckelino, the fermionic
degree of freedom of the \Stuckelberg multiplet, is the lightest
supersymmetric particle (LSP). We compute the relic density of
this particle including coannihilations with the next to lightest
supersymmetric particle (NLSP) and with the next to next to
lightest supersymmetric particle (NNLSP) which are assumed almost
degenerate in mass. This assumption is needed in order to satisfy
the stringent limits that the Wilkinson Microwave Anisotropy Probe
(WMAP) puts on the relic density. We find that the WMAP
constraints can be satisifed by different NLSP and NNLSP
configurations as a function of the mass gap with the LSP. These
results hold in the parameter space region where the model remains
perturbative.
\end{abstract}

\end{titlepage}

%\tableofcontents

\section{Introduction}

A great deal of work has been done recently to embed the standard
model of particle physics (SM) into a brane construction
\cite{Marchesano:2007de,Blumenhagen:2005mu,Lust:2004ks,Kiritsis:2003mc}.
This research is part of the effort, initiated in
\cite{Candelas:1985en}, to build a fully realistic four
dimensional vacuum out of string theory. While the original models
were formulated in the framework of the heterotic string, the most
recent efforts were formulated for type II strings in order to
take advantage of the recent work on moduli stabilization using
fluxes. Such brane constructions naturally lead to extra anomalous
$U(1)$'s in the four dimensional low-energy theory and, in turn,
to the presence of possible heavy $Z^\prime$ particles in the
spectrum. These particles should be among the early findings of
LHC and besides for the above cited models they are also a
prediction of many other theoretical models of the unification of
forces (see \cite{Langacker:2008yv} for a recent review). It is
then of some interest to know if these $Z^\prime$ particles
contribute to the cancelation of the gauge anomaly in the way
predicted from string theory or not. In \cite{ourpaper} some of
the present authors have studied a supersymmetric (SUSY) extension
of the minimal supersymmetric standard model (MSSM) in which the
anomaly is canceled {\it \`a la} Green-Schwarz. The model is only
string-inspired and is not the low-energy sector of some brane
construction. The reason of this choice rests in our curiosity to
explore the phenomenology of these models keeping a high degree of
flexibility, while avoiding the intricacies and uncertainties
connected with a string theory construction. For previous work
along these lines we refer to
\cite{Anastasopoulos:2006cz}-\cite{Klein:1999im}. In this work we
perform a consistency check of our model \cite{ourpaper} by
evaluating the thermal relic density and comparing it against the
WMAP data.

The cancelation of the $U(1)'$ anomaly in our model requires the
introduction of an extra complex scalar field whose supersymmetric
partner is called the St\"uckelino. We will see  in the following
that if the  latter is the lightest supersymmetric particle (LSP)
its interactions are such to define it as an XWIMP (i.e. a weakly
interacting particle with couplings at least one order of
magnitude less than the standard weak interactions): in fact it
will also turn out to be a cold relic. If the St\"uckelino is the
lightest supersymmetric particle (LSP), its relic density turns
out to be too high with respect to the experimental data. This is
why, following \cite{Feldman:2006wd}, we favor a next to lightest
supersymmetric particle (NLSP) with a mass close to the LSP. We
show that an interesting scenario arises also for three particle
coannihilation processes. In these two cases, the model is
consistent with the experimental data. Moreover in the three
particle case we find configurations in which the LSP and the NLSP
do not need to be nearly degenerate in mass. In this case the mass
gap between the two can be of the order of $20\%$. This is the
plan of the paper: in Section 2 we describe our model. In Section
3 and 4 we find the LSP and study the St\"uckelino interactions.
Finally in Section 5 we compute the relic density. Section 6 is a
summary of our results.

\section{Model Setup} \label{sect:Model Setup}

In this section we briefly discuss our theoretical framework. We
assume an extension of the MSSM with an additional abelian vector
multiplet $V^{(0)}$ with arbitrary charges. The anomalies are
canceled with the Green-Schwarz (GS) mechanism and with the
Generalized Chern-Simons (GCS) terms. All the details can be found
in~\cite{ourpaper}. All the MSSM fields are charged under the
additional vector multiplet $V^{(0)}$, with charges which are
given in Table~\ref{QTable}, where $Q_i, L_i$ are the left handed
quarks and leptons respectively while $U^c_i, D^c_i, E^c_i$ are
the right handed up and down quarks and the electrically charged
leptons. The superscript $c$ stands for charge conjugation. The
index $i=1,2,3$ denotes the three different families. $H_{u,d}$
are the two Higgs scalars.
  \begin{table}[h]
  \centering
  \begin{tabular}[h]{|c|c|c|c|c|}
   \hline & SU(3)$_c$ & SU(2)$_L$  & U(1)$_Y$ & ~U(1)$^{\prime}~$\\
   \hline $Q_i$   & $\bth$       &  $\btw$       &  $1/6$   & $Q_{Q}$ \\
   \hline $U^c_i$   & $\bar \bth$  &  $\bon$       &  $-2/3$  & $Q_{U^c}$
\\
   \hline $D^c_i$   & $\bar \bth$  &  $\bon$       &  $1/3$   & $Q_{D^c}$
\\
   \hline $L_i$   & $\bon$       &  $\btw$       &  $-1/2$  & $Q_{L}$ \\
   \hline $E^c_i$   & $\bon$       &  $\bon$       &  $1$     &
$Q_{E^c}$\\
   \hline $H_u$ & $\bon$       &  $\btw$       &  $1/2$   & $Q_{H_u}$\\
   \hline $H_d$ & $\bon$       &  $\btw$       &  $-1/2$  & $Q_{H_d}$ \\
   \hline
  \end{tabular}
  \caption{Charge assignment.}\label{QTable}
  \end{table}

The key feature of this model is the mechanism of anomaly
cancelation. As it is well known, the MSSM is anomaly free. In our
MSSM extension all the anomalies that involve only the $SU(3)$,
$SU(2)$ and $U(1)_Y$ factors vanish identically. However,
triangles with $U(1)'$ in the external legs in general are
potentially anomalous. These anomalies are\footnote{We are working
in an effective field theory framework and we ignore throughout
the paper all the gravitational effects.  In particular, we do not
consider the gravitational anomalies which, however, could be
canceled by the Green-Schwarz mechanism.}
\bea
   U(1)'-U(1)'-U(1)':     &&\ \cA^{(0)}  \\
   U(1)'-U(1)_Y - U(1)_Y: &&\ \cA^{(1)} \\
   U(1)'-SU(2)-SU(2):     &&\ \cA^{(2)} \\
   U(1)'-SU(3)-SU(3):     &&\ \cA^{(3)} \\
   U(1)'-U(1)'-U(1)_Y:    &&\ \cA^{(4)}
\eea
All the remaining anomalies that involve $U(1)'$s vanish
identically due to group theoretical arguments (see Chapter 22 of
\cite{Weinberg2}). Consistency of the model is achieved by the
contribution of a \Stuckelberg field $S$ and its appropriate
couplings to the anomalous $U(1)'$. The \Stuckelberg lagrangian
written in terms of superfields is \cite{Klein:1999im}
  \be
 \L_S = {\frac{1}{4}} \left. \( S + S^\dagger + 4 b_3 V^{(0)} \)^2
\right|_{\thth}\!\!
                - {\frac{1}{2}} \left\{ \[\sum_{a=0}^2 b^{(a)}_2 S
\Tr\( W^{(a)}
W^{(a)} \) + b^{(4)}_2 S W^{(1)} W^{(0)} \]_{\th^2} \!\!\!\!\!\!+h.c.\!
\right\}
   \label{Laxion}
\ee
where the index $a=0,\ldots,3$ runs over the $U(1)',\, U(1)_Y,\,
SU(2)$ and $SU(3)$ gauge groups respectively. The \Stuckelberg
multiplet is a chiral superfield
 \be
    S =  s+ i\sqrt2 \th \psi_S + \th^2 F_S - i \th \s^\m \bar\th \pd_\m s
+
                {\frac{\sqrt2}{2}}  \th^2 \bar\th \bar\s^\m \pd_\m \psi_S
-
{\frac{1}{4}} \thth \Box s \label{Smult}
 \ee
The lowest component of $S$ is a complex scalar field $s=\a+i\f$.
In our scenario the scalar $\f$ is eaten up in the \Stuckelberg
mechanism to give mass to the gauge field. On the other end, the
scalar $\a$ is the dilaton of string theory and its value must be
determined somehow. We will not investigate the way in which this
happens, but we will just retain the final result: $\a$  drops out
of our effective lagrangian. In the string literature (see for
instance \cite{ourpaper}-\cite{Feldman:2006wd}) the fields $\f$,
$\a$ and $\psi_S$ are respectively known as axion, saxion and
axino. In this paper, to avoid confusion with the much better
known QCD axion/axino system (see for instance
\cite{Rajagopal:1990yx}-\cite{axinoCDM}), we will adopt the
convention of \cite{Kors:2004iz} and, from now on, we will refer
to $s$ as the \Stuckelberg scalar and to $\psi_S$ as the
St\"uckelino.

The \Stuckelberg multiplet $S$ transforms under $U(1)'$ as
\bea
   V^{(0)} &\to& V^{(0)} + i \( \Lambda - \Lambda^\dag \) \nn\\
   S  &\to& S - 4 i ~b_3 ~\Lambda \label{U1'}
  \label{U1Trans}\eea
where $b_3$ is a constant related to the $Z'$ mass. In our model
there are two mechanisms that give mass to the gauge bosons: (i)
the \Stuckelberg mechanism and (ii) the Higgs mechanism. In this
extension of the MSSM, the mass terms for the gauge fields for
$\QHu=-\QHd=0$\footnote{We impose this condition to simplify our
computations and to give analytical expressions of limited
dimensions. There are no obstructions to set  $\QHu=-\QHd\neq 0$.}
are given by
     \be
      \L_M = \frac{1}{2} \(V^{(0)}_\m \ V^{(1)}_\m \ V^{(2)}_{3\m} \) M^2
                          \( \begin{array}{c} V^{(0)\m}\\ V^{(1)\m}\\
V^{(2)\m}_{3}
\end{array} \)
     \ee
with $M^2$ being the gauge boson mass matrix
     \be
      M^2= \( \begin{array}{ccc} M_{V^{(0)}} & ~~~0& ~~~0  \\
                 ... & g_1^2 \frac{v^2}{4} & -g_1 g_2 \frac{v^2}{4}   \\
                 ... & ...  & g_2^2 \frac{v^2}{4}  \\\end{array} \)
  \label{BosonMasses} \ee
where $M_{V^{(0)}}=4 b_3 g_0$  is the mass parameter for the anomalous
$U(1)$ and it
is assumed to be in the TeV
range. The lower dots denote the obvious terms under symmetrization.
After diagonalization, we obtain the eigenstates
 \bea
   A_\m &=&\frac{g_2 V^{(1)}_\m + g_1 V^{(2)}_{3\m}}{\sqrt{g_1^2+g_2^2}}
\label{photon}\\
    Z_{0\m} &=& \frac{g_2 V^{(2)}_{3\m} - g_1
V^{(1)}_\m}{\sqrt{g_1^2+g_2^2}}
\label{Z0}\\
   Z'_\m  &=& V^{(0)}_\m \label{Zprime}
    \eea
and the corresponding masses
   \bea
   M^2_{\g}&=&0\\
    M^2_{Z_0} &=&\frac{1}{4} \(g_1^2+g_2^2\) v^2\label{Z0mass}\\
M^2_{Z'}  &=&M_{V^{(0)}}^2
  \label{Zpmass} \eea
Finally the rotation matrix from the hypercharge to the photon
basis is
    \bea
     \( \begin{array}{c} Z'_\m\\
                         Z_{0 \m}\\
                         A_\m \end{array} \)
       &=&O_{ij}
     \( \begin{array}{c} V^{(0)}_\m\\
                         V^{(1)}_\m\\
                         V^{(2)}_{3\m} \end{array} \) \label{Oij}
      =\( \begin{array}{ccc}      1& 0  ~~& 0  \\
                                0& - \sin\theta_W& \cos\theta_W  \\
                                0 & \cos\theta_W & \sin\theta_W  \\
           \end{array} \)
     \( \begin{array}{c} V^{(0)}_\m\\
                         V^{(1)}_\m\\
                         V^{(2)}_{3\m} \end{array} \) \eea
where $i,j=0,1,2$.

We now give the expansion of the lagrangian piece $\L_S$ defined in (\ref{Laxion}) in component fields only for the part that
is needed in the following sections. Using the Wess-Zumino gauge we get
\bea
 \L_{\text{St\"uckelino}}&=&{\frac{i}{4}} \psi_S \s^\m \pd_\m \psib_S -\sqrt2 b_3 \psi_S
\l^{(0)}
   -{\frac{i}{2\sqrt2}} \sum_{a=0}^2b^{(a)}_2  \Tr \( \l^{(a)} \s^\m
\sb^\n F_{\m
\n}^{(a)} \) \psi_S \nn\\
   &&-{\frac{i}{2\sqrt2}} b^{(4)}_2 \[ {\frac{1}{2}} \l^{(1)} \s^\m \sb^\n
F_{\m
\n}^{(0)} \psi_S
   + (0 \leftrightarrow 1) \] +h.c.
\label{axinolagr}\eea
As it was pointed out in \cite{Anastasopoulos:2006cz}, the
\Stuckelberg mechanism is not enough to cancel all the anomalies.
Mixed anomalies between anomalous and non-anomalous factors
require an additional mechanism to ensure consistency of the
model: non-gauge invariant GCS terms must be added. In our case,
the GCS terms have the form \cite{Andrianopoli:2004sv}
   \bea
 \L_{GCS} &=&- d_4     \[ \( V^{(1)} D^\a V^{(0)} - V^{(0)} D^\a V^{(1)}\) W^{(0)}_\a + h.c. \]_{\thth} +\nn\\
        &&+  d_5     \[ \( V^{(1)} D^\a V^{(0)} - V^{(0)} D^\a V^{(1)}\) W^{(1)}_\a + h.c. \]_{\thth} +\nn\\
       &&+  d_6 \Tr \bigg[ \( V^{(2)} D^\a V^{(0)} - V^{(0)} D^\a V^{(2)}\) W^{(2)}_\a + n.a.c + h.c. \bigg]_{\thth}
   \label{GCS_1}
\eea
where $n.a.c.$ refers to non-abelian completion terms. The $b$
constants in (\ref{Laxion}) and the $d$ constants in (\ref{GCS_1})
are fixed by the anomaly cancelation procedure (for details
see~\cite{ourpaper}).

For a symmetric distribution of the anomaly, we have
\bea
 &&     b^{(0)}_2 b_3 =-\frac{\cA^{(0)}}{384\pi^2}~
 \qquad b^{(1)}_2 b_3 = - \frac{\cA^{(1)}}{128 \pi^2}
 \qquad b^{(2)}_2 b_3 = -\frac{\cA^{(2)}}{64 \pi^2}~
 \qquad b^{(4)}_2 b_3 = -\frac{\cA^{(4)}}{128 \pi^2}\nn\\
 &&~~~~ d_4 =- \frac{\cA^{(4)}}{384 \pi^2}~~~~~
 \qquad d_5 = \frac{\cA^{(1)}}{192\pi^2}~~~~~~
 \qquad d_6= \frac{\cA^{(2)}}{96 \pi^2}
\label{bsds}  \eea
It is worth noting that the GCS coefficients $d_{4,5,6}$ are fully
determined in terms of the $\cA$'s by the gauge invariance, while the
$b_2^{(a)}$'s
depend
only on the free parameter $b_3$, which is related to the mass of
the anomalous $U(1)$.

The soft breaking sector of the model is given by
   \be
    \L_{soft}=\L_{soft}^{MSSM}- {\frac{1}{2}}  \(M_0 \l^{(0)} \l^{(0)} + h.c. \)
    - {\frac{1}{2}}  \(\frac{M_S}{2} \psi_S \psi_S  + h.c. \) \label{Lsoft}
   \ee
where $\L_{soft}^{MSSM}$ is the usual soft susy breaking lagrangian while
 $\l^{(0)}$ is the gaugino of the added
$U(1)'$ and $\psi_S$ is the St\"uckelino. The St\"uckelino soft
mass term deserves some comment: from~\cite{Girardello:1981wz} we
know that a fermionic mass term for a chiral multiplet is not
allowed in presence of Yukawa interactions in which this chiral
multiplet is involved. But in the classical Lagrangian the
\Stuckelberg multiplet cannot contribute to superpotential terms
given that the gauge invariance given from our U(1)' symmetry
(\ref{U1Trans}) requires non-holomorphicity in the chiral fields.
In fact in our model both the St\"uckelino and the scalar $\f$
couple only through GS interactions. It is worth noting that a
mass term for the scalar $\f$ is instead not allowed since it
transforms non trivially under the anomalous $U(1)'$ gauge
transformation~(\ref{U1Trans}).

\section{Neutralino Sector \label{Neutralinos}}
Assuming the conservation of R-parity the LSP is a good weak interacting massive particle (WIMP) dark matter candidate.
As in the MSSM the LSP is given by a linear combination of fields in the
neutralino sector.
The general form of the neutralino mass matrix is given
in~\cite{ourpaper}. Written
in the interaction eigenstate basis
$(\psi^{0})^T= (\psi_S, \ \l^{(0)},\ \l^{(1)} ,\ \l^{(2)}_3,\ \tilde h_d^0,\ \tilde
h_u^0)$
it is a six-by-six matrix.
From the point of view of the strength of the interactions the two extra
 states are
not on the same footing with respect to the standard ones. The
St\"uckelino and the extra gaugino $\l^{(0)}$ dubbed primeino are
in fact extremely weak interacting massive particle (XWIMP). Thus
we are interested in situations in which the extremely weak sector
is decoupled from the standard one and the LSP belongs to this
sector. This can be achieved at tree level with the choice
\be
 \QHu=\QHd=0 \label{QHuchoice}
\ee
The neutralino mass matrix $ {\bf M}_{\tilde N} $ becomes
   \be
    {\bf M}_{\tilde N}
     =   \(\begin{array}{cccccc}
          \frac{M_S}{2} & \frac{M_{V^{(0)}}}{\sqrt2} &   0   &   0   &
     0
    & 0 \\
                  \dots &       M_0                  &   0   &   0   &
     0
    & 0   \\
                  \dots &      \dots                 &  M_1  &   0   &
-\frac{g_1
v_d}{2}& \frac{g_1 v_u}{2} \\
                  \dots &      \dots                 & \dots &  M_2  &
\frac{g_2
v_d}{2} & -\frac{g_2 v_u}{2} \\
                  \dots &      \dots                 & \dots & \dots &
     0
    & -\m  \\
                  \dots &      \dots                 & \dots & \dots &
   \dots
    & 0
         \end{array}\) \label{massmatrix} ~~~~
   \ee
where $M_S,~M_0,~M_1,~M_2$ are the soft masses coming from the soft
breaking terms
(\ref{Lsoft}) while $M_{V^{(0)}}$ is given in
~(\ref{BosonMasses}).
It is worth noting that the D terms and kinetic mixing terms
can be neglected in the tree-level computations of the eigenvalues and
eigenstates.

\begin{figure}[t]
\begin{center}
\includegraphics[scale=0.4]{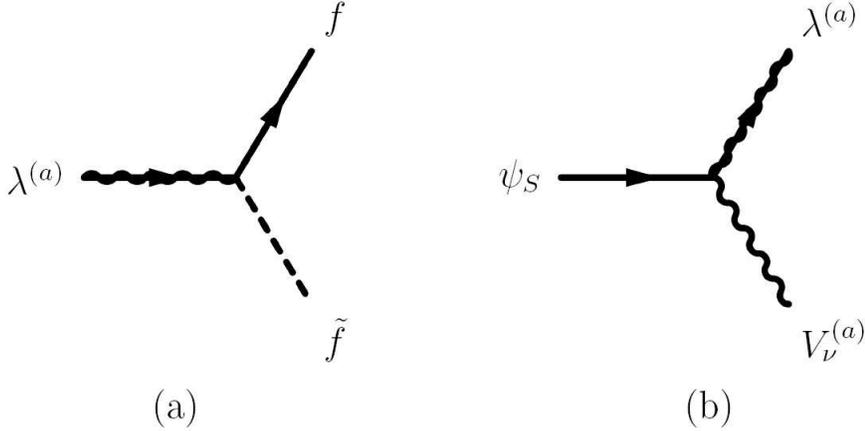}
  \caption{(a) Gaugino-fermion-sfermion interaction vertex. (b) St\"uckelino-gaugino-vector interaction vertex.}
\label{Upint}
\end{center}
\end{figure}

Moreover, we make the assumption that
\be \
M_0\gg M_S,M_{V^{(0)}}
\label{axinodecoupling}
\ee
so that the St\"uckelino is the LSP. This assumption is motivated
by the interaction strengths of the two extra states: the
St\"uckelino interacts via the vertex shown in Fig.~\ref{Upint}b
which can easily be seen (from (\ref{axinolagr})) to be
proportional to the coefficient $ b^{(a)}_2$ of (\ref{bsds})
which, in turn, is inversely proportional to $b_3$ that is the $
Z'$ mass given that $M_{ Z'}=M_{V^{(0)}}=4 b_3 g_0$. Given these
considerations the vertex shown in Fig.~\ref{Upint}b is then of
order $\sim g_0 g_a^2/\MZp$. The primeino interacts via the vertex
in Fig.~\ref{Upint}a which is of order $\sim g_a$ that is the
standard strength of weak interactions. Assuming the two
decoupling relations~(\ref{QHuchoice}) and~(\ref{axinodecoupling})
we will see in the following sections that
 the dominant contribution in the (co)annihilation processes is that of the primeino, which is of the type of a standard gaugino interaction.
\section{St\"uckelino Interactions \label{Axino Interactions}}
 The St\"uckelino interactions can be read off from the interaction
lagrangian~(\ref{axinolagr}).  The relevant
St\"uckelino-MSSM neutralino interaction term, written in terms
of four components Majorana spinors\footnote{The gamma matrices
$\gamma^\mu$ are in the Weyl representation. We use
capital letters for four components spinors and lower case letters
for two components spinors. }, is given by: \be
 \L =i \sqrt{2} g_1^2 b_2^{(1)}  \bar\Lambda^{(1)} \g_5 [\g^\m,\g^\n](\partial_\m V^{(1)}_\n)    \Psi_S +
        i {\frac{\sqrt{2}}{2}} g_2^2 b_2^{(2)}  \bar\Lambda^{(2)}_3 \g_5 [\g^\m,\g^\n](\partial_\m V^{(2)}_{3\n}) \Psi_S
\label{Lint4d}
\ee
where the $b_2^{(a)}$
coefficients are given in~(\ref{bsds}). The related interaction vertex Feynman rule is
\be
C^{(a)} \gamma_5 [\gamma^{\mu}, \gamma^{\nu}]i k_{\mu}
 \label{lpg}
 \ee
where $k_\m$ is the momentum of the outgoing vector and the $C^{(a)}$'s are
\bea
 C^{(1)}=\sqrt{2}g_1^2 b_2^{(1)} \nn\\
 C^{(2)}={\frac{\sqrt{2}}{2}} g_2^2 b_2^{(2)}
 \label{vertice}
\eea
The interaction Lagrangian (\ref{Lint4d}) expressed in the mass eigenstates basis is
\bea
 \L &=& i \sum_j \tilde{N_j} \(  \cos\theta_W  C^{(1)}  N_{(1)j} + \sin\theta_W  C^{(2)}  N_{(2)j} \) \g_5 [\g^\m,\g^\n](\partial_\m A_\n)    \Psi_S + \nn\\
     && i \sum_j \tilde{N_j} \( -\sin\theta_W  C^{(1)}  N_{(1)j} + \cos\theta_W  C^{(2)}  N_{(2)j} \) \g_5 [\g^\m,\g^\n](\partial_\m {Z_0}_\n)    \Psi_S
\label{Lintphys}
\eea
where $\tilde{N_j}$ is a generic neutralino and $N_{ij}$ is the
matrix that diagonalizes (\ref{massmatrix}). We remind that the
St\"uckelino $\Psi_S$ is directly a mass eigenstate because of the
decoupling (\ref{axinodecoupling}). We stress that these are only
interactions between the St\"uckelino and the MSSM neutralinos.
All the Stuckelino interactions in (\ref{axinolagr}) include also
analogous interactions involving the charged wino or the primeino.

\begin{figure}[t]
\begin{center}
\includegraphics[scale=0.4]{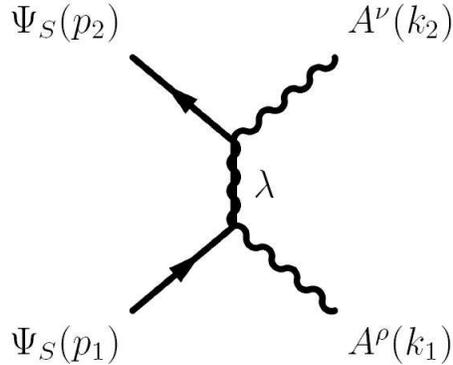}
      \caption{Annihilation of two St\"uckelinos into two gauge vectors via the
exchange of a gaugino.}
      \label{figura1}
\end{center}
\end{figure}

\begin{figure}[t]
\begin{center}
\includegraphics[scale=0.4]{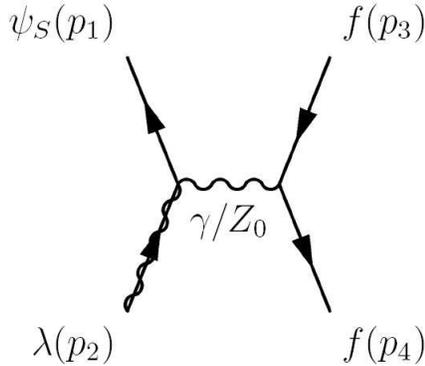}
      \caption{Coannihilation of a St\"uckelino and a bino into a $f \bar f$
pair via the exchange of a photon or a $Z_0$.}
      \label{figuracoann}
\end{center}
\end{figure}

The factors in~(\ref{vertice}) are of naive dimension higher than
four. They then contain the parameters $b_2^{(a)}$ which are
inversely proportional to the mass of the $Z'$ (see
~(\ref{bsds})). Given these  interactions, our St\"uckelino will
be an extremely weak interacting particle, according to the
definition we gave in the introduction. The simplifying assumption
(\ref{QHuchoice}) has now decoupled our St\"uckelino and primeino
from the standard MSSM sectors and, at tree level, the relevant
diagrams are given in Figs.~\ref{figura1} and ~\ref{figuracoann}. We can now give a rough estimate of the
effective interaction, comparing with the standard Fermi coupling,
$G_F$, of weak interactions: given a mass of the $Z'$ boson of the
order of the TeV, the effective coupling in  Fig.~\ref{figura1} is
$G_F'\approx 10^{-4} G_F$ and that in
 Fig.~\ref{figuracoann} is $G_F'\approx 10^{-2} G_F$.
Let us go back now to Fig.~\ref{figura1}, where we denoted with
$p_1$ and $p_2$ the incoming momenta of the St\"uckelinos while
$k_1$ and $k_2$ are the two outcoming momenta of the gauge bosons
in the final state. We will concentrate on the case with two
photons in the final state. In this case the result for the
differential cross section is given by \be \
\frac{d\s}{d\Omega}=\frac{4M_S^2 \omega_1}{16 \pi^2
(\omega_1+\omega_2)^2(\sqrt{M_S^2-E_2^2})}\sum_{i,j=1}^2\mathcal{M}_i
\mathcal{M}^*_j \label{axinoaxinodiffcs} \ee where $\omega_1$ and
$\omega_2$ are the energies of the two outcoming photons. Each
amplitude $\mathcal{M}_i$ is proportional to the relative
coefficient $(C^{(a)})^2$ whose generic form is given
in~(\ref{vertice}). In our scenario the St\"uckelino annihilations
alone, being the cross section~(\ref{axinoaxinodiffcs}) extremely
weak, cannot give a relic density in the WMAP preferred range.
Thus, in the scenario of an XWIMP St\"uckelino, we are forced to
consider coannihilations between the St\"uckelino and the NLSP
\cite{Feldman:2006wd}. Several scenarios can be considered for the
NLSP. We can split them into two major classes: one in which the
NLSP is either a pure bino or a pure wino, and thus a
coannihilation with a third MSSM particle is needed in order to
recover the WMAP result, and one in which the NLSP is a generic
MSSM neutralino with a non-negligible bino and/or wino component.
In both classes in order to have effective coannihilations, the
NLSP (and eventually the other MSSM particle involved in the
coannihilation process) must be almost degenerate in mass.
Furthermore, in the most common applications, the cross sections
for the annihilations of the the LSP and the NLSP and that of the
coannihilations between the LSP and NSLP are roughly of the same
order of magnitude. In our case this last condition will be
stretched and the cross section we are going to discuss will
differ for some orders of magnitude. This situation is not
completely new in literature: already in \cite{reliccoann} in
which the NLSP is the stop these differences are of order
$10^{-2}$. In \cite{Feldman:2006wd} differences of order $\leq
10^{-4}$ are considered, while in \cite{Edsjo:2003us} $10^{-4}\div
10^{-5}$ differences are found\footnote{This can be extracted from
Fig.4 of the previous reference after an appropriate rescaling.}.
But what assures us that the two species are still in thermal
equilibrium and do not decouple separately? The existence of
interactions of the type $LSP+MSSM1\to NLSP+MSSM2$ is then
required to keep the LSP in equilibrium \cite{Edsjo:1997bg}.
$MSSM1, MSSM2$ are two MSSM particles. Furthermore $MSSM1$ better
be relativistic so that its abundance is much larger of any cold
particle to foster the above reaction. The above reaction will
then keep the LSP at equilibrium and the formalism of
coannihilation can be safely employed. As we will see in the next
section, all of these requirements are met in our scenario.

As a first example we then consider a pure bino as the NLSP. The
allowed coannihilation processes with the St\"uckelino are those
which involve an exchange of a photon or a $Z_0$ in the
intermediate state and with a SM fermion-antifermion pair, Higgses
and $W$'s in the final state. The diagram with the
fermion-antifermion in the final state is sketched in
Fig.~\ref{figuracoann}. The differential cross section in the
center of mass frame has the following general form \be
 \frac{d \s}{d \Omega} \propto \frac{1}{s}
\frac{p_f}{p_i} |\M|^2
\ee
where $s$ is the usual Mandelstam variable and $p_{f,i}$ is the spatial
momentum of the outgoing (incoming) particles.
On dimensional ground $ |\M|^2$ has at least a linear dependence on $p_f$
and this
implies that the dominant contribution comes from the diagram with the SM
fermion-antifermion pair $f$ and $\bar{f}$ in the final state:
\be
\Psi_S \l^{(a)}\rightarrow f \bar{f}
\ee
The resulting differential cross section, computed in the center of mass
frame, is
\be
\frac{d\s}{d \Omega}=\sum_{f} c_f
\frac{\sqrt{(E_3-m_{f})^2}}{64 \pi^2 (E_1+E_2)^2
\sqrt{(E_1^2-M_{S}^2)}}(\M_\g^2+\M_{Z_0}^2+\M^*_\g \M_{Z_0}+\M_\g
\M_{Z_0}^*)
\label{sezcoan}
\ee
where the sum is extended to all the SM fermions (with mass $m_f$) while
$c_f$ is a color factor. Details of the amplitude computation can be found in Appendix~\ref{nnAmp}.

\section{St\"uckelino Relic Density} \label{sect:Axino Relic Density}
In this section we compute the relic density of the St\"uckelino.
The case of the St\"uckelino as a cold dark matter candidate has
been studied for the first time in~\cite{axinoCDM}. As we said in
the previous section we study two scenarios: the first in which
 the St\"uckelino coannihilates with only one NLSP degenerate in mass (a generic MSSM neutralino), the second in which there is an additional
supersymmetric particle (either a chargino or a stau) involved in
the coannihilation process with the St\"uckelino and the NLSP. In
the following we will be largely following \cite{Feldman:2006wd}.
Since this is a first study and given also the simplifying choice
(\ref{QHuchoice}) we will defer a complete analysis to a future
work and will content ourselves with showing that our model can
accommodate for WMAP data. Then, following this philosophy we will
not solve the Boltzmann equation numerically but, in agreement
with \cite{Feldman:2006wd}, we will argue that, if the ratio
between the thermally averaged cross sections of the
coannihilation of Fig.~\ref{figuracoann} and that of a typical
neutralino annihilation is much less than one, a relic density
satisfying the WMAP requirements can be found.

Just to fix the notation we briefly review the relic density
computation for $N$ interacting
species~\cite{reliccoann,Edsjo:2003us,Edsjo:1997bg}. The
requirements discussed in the previous section are met by our
model given the lagrangian (\ref{axinolagr}) and the condition of
(\ref{QHuchoice}). In this case all channels are open to
interactions and the St\"uckelino has an interaction with the
photon and the bino of strength $b_2^{(1)}$\footnote{There is
another interesting possibility if we insist on imposing
(\ref{QHuchoice}): taking also $Q_L=0$ SUSY terms of the type
$\int d^2\theta\int d^2\bar\theta
\log(S+S^\dagger+V^{(0)})/M_S(LH_u/M_S^2+H_u^\dagger
L^\dagger/M_S^2)$ can be safely added to the lagrangian. This term
induces a vertex between the St\"uckelino, neutrino and Higgs
field which can be a viable candidate to keep the St\"uckelino in
thermal equilibrium.}.

The Boltzmann equation for $N$ particle species is given by:
\be
\frac{dn}{dt}=-3Hn-\sum_{i,j=1}^N \langle \s_{ij} v_{ij} \rangle
(n_i n_j - n^{eq}_i n^{eq}_j) \label{boltz3}
\ee
where $n_i$ denotes the number density per unit of comoving volume of the
species $i=1,\ldots,N$ ($i=1$ refers to the LSP, $i=2$ refers to the NLSP, and so on), $n=\sum_i
n_i$, $H$ is
the Hubble constant, $\s_{ij}$ is the annihilation cross section between a
species
$i$ and a species $j$, $ v_{ij}$ is the modulus of the relative velocity
while
$n_i^{eq}$ is the equilibrium number density of the species $i$ given by:
\begin{equation}
 \frac{ n_i^{eq} }{ n^{eq} }=\frac{ g_i\left(1+\Delta_i\right)^{3/2} e^{-\Delta_i x_f} }{\sum_i g_i\left(1+\Delta_i\right)^{3/2} e^{-\Delta_i x_f}}
\end{equation}
where $g_i$ are the internal degrees of freedom, $\Delta_i=(m_i-m_1)/m_1$.
$x_f=m_1/T$ is known once (\ref{boltz3}) is solved.
A preliminary estimate of $x_f$ can be obtained by solving the equation
\begin{equation}
x\simeq \ln \left(x^{1/2}M_P\, m_{LSP} \left<\sigma
v\right>\right) \label{xf}\end{equation} which can be obtained
from the decoupling condition. This can be done using the
estimate\footnote{To check this crude estimate in the case of our
St\"uckelino, we have also solved numerically $\left<\sigma
v\right>$ sweeping the temperatures range $1\div 100$ GeV. We have
then fit the results to recover a function in good agreement with
(\ref{decoupleq}).}
\begin{equation}
\left<\sigma v\right>\simeq G^2\, m_{LSP}^2 x^{-5/2}
\label{decoupleq}
\end{equation}
where the effective coupling $G$ can be the $G_F, G_F'$ introduced
in Section \ref{Axino Interactions}. In the mass range
$m_{LSP}=10\div 1000$ GeV, by plugging in (\ref{decoupleq})
$G=G_F$ we would get $x_f=25\div 30$ (if we would take our
St\"uckelino as a separate species, that is we would use $G=G_F'$,
we would get those values divided by half).

Eq.~(\ref{boltz3}) can be rewritten in a useful way by defining the thermal average of the effective
cross section

\be \langle \s_{eff} v \rangle \equiv \sum_{i,j=1}^N\langle
\s_{ij} v_{ij} \rangle \frac{n_i^{eq}}{n^{eq}}
\frac{n_j^{eq}}{n^{eq}} \label{sigmaeff} \ee obtaining \be
\frac{dn}{dt}=-3Hn-\langle \s_{eff} v \rangle (n^2 - (n^{eq})^2)
\label{Boltz} \ee where $n^{eq}=\sum_i n_i^{eq}$. It is sensible
to use these approximations when the LSP is kept in equilibrium by
a relativistic particle in the thermal background as discussed in
the previous section. Given the typical values of $x_f$ discussed
above, the ratio between the number density per comoving volume of
this relativistic species and that of a cold relic is
$n^{eq}_{rel}/n^{eq}_{cold}=10^4\div 10^{6}$, which is sufficient
to keep the St\"uckelino coupled until the end of coannihilations.

As a rule of thumb~\cite{Jungman:1995df} a first order estimate of the relic density is given by
\begin{equation}
 \Omega_\chi h^2\simeq \frac{10^{-27}\,{\rm cm^3\, s^{-1}}}{\langle \s_{eff} v \rangle}
\label{oh2vssigmaeff}
\end{equation}

\begin{figure}[t]
      \begin{center}
\raisebox{0.7ex}[0cm][0cm]{\unitlength=1.mm
      \includegraphics[scale=0.28]{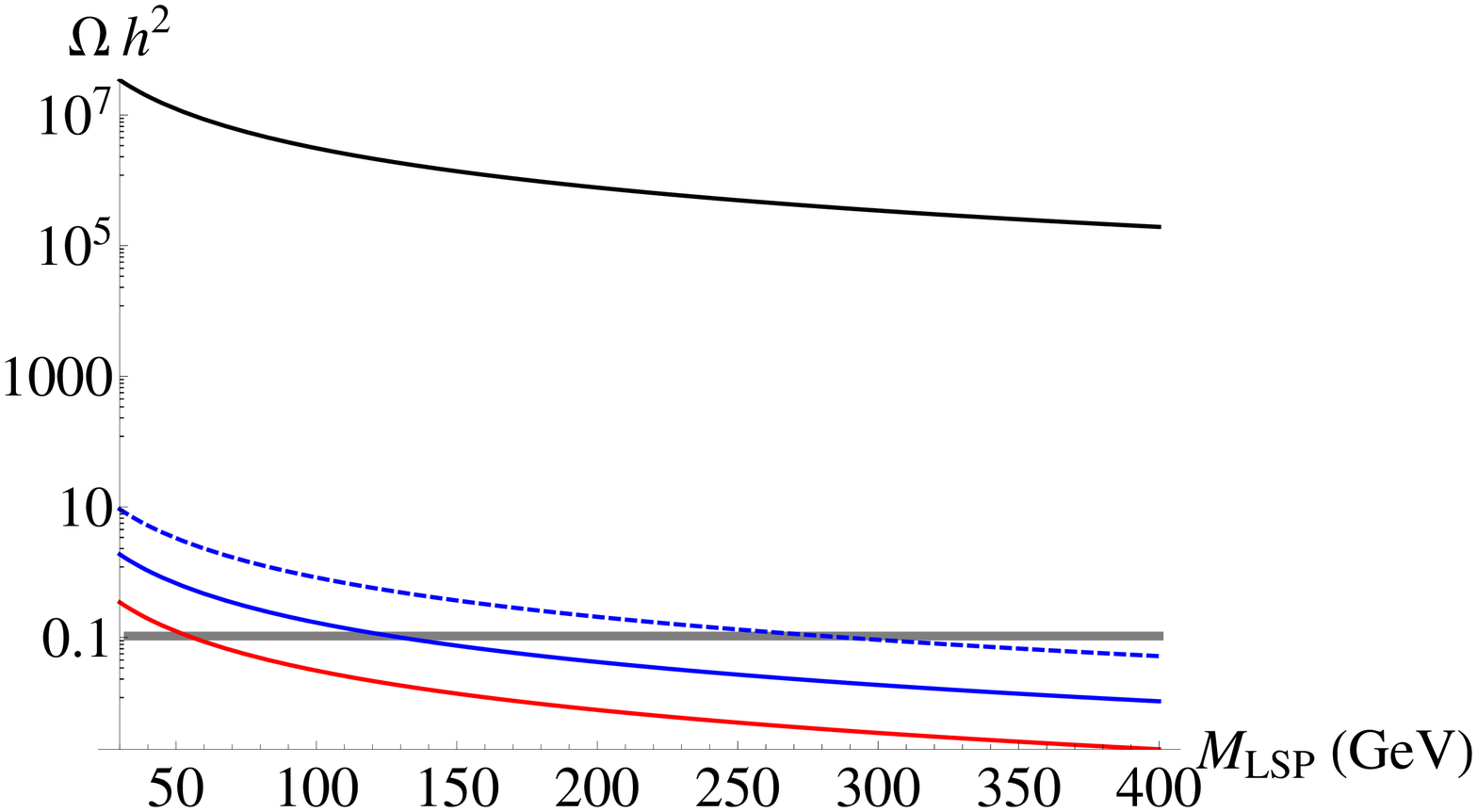}}
      \includegraphics[scale=0.32]{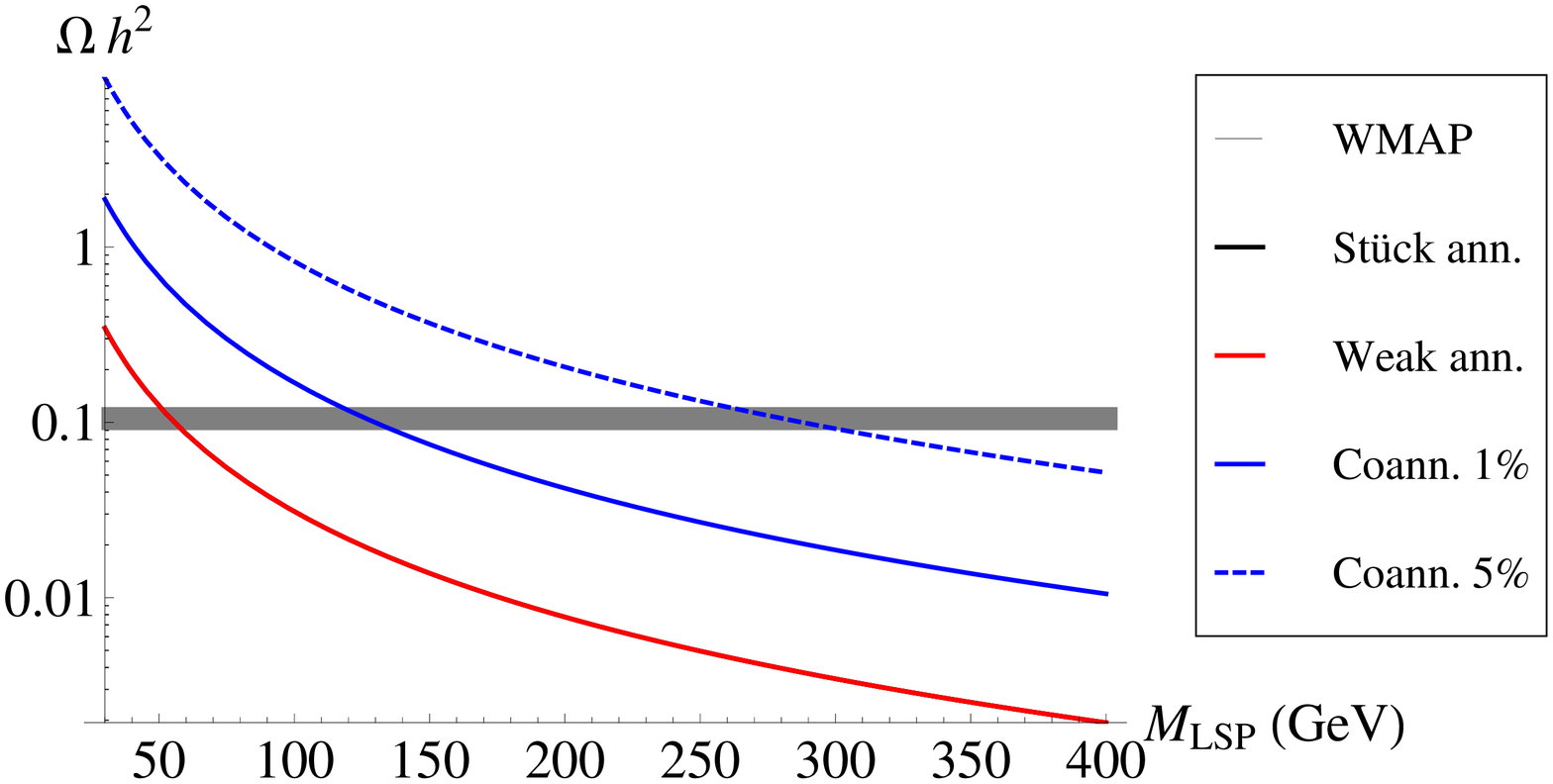}
      \caption{Relic density estimation for a pure St\"uckelino cross section (black), a pure electro-weak cross section (red),
               a coannihilating cross section with a St\"uckelino-neutralino mass gap $\D=1\%$ (blue)
               and a coannihilating cross section with $\D=5\%$ (dashed blue). The  WMAP data range is the gray strip. The Figure on the right
is a zoom of that on the left around the region of interest.}
\label{DMestimate}
\end{center}
\end{figure}

To give a rough idea of the role played by
coannihilations we plotted in Fig. \ref{DMestimate} the relic
density estimate (\ref{oh2vssigmaeff}) induced by an electro-weak
cross section ((\ref{decoupleq}) with $G=G_F$), a St\"uckelino
cross section ((\ref{decoupleq}) with $G=G'_F$) and two
coannihilations cross sections estimations. We can see that the
St\"uckelino annihilations cannot give a relic density in the WMAP
data range, while coannihilations can do it. Moreover we can see
that the effect of coannihilations on the MSSM is to decrease the
efficiency of the MSSM annihilations (while they increase the
St\"uckelino one) and to increase the LSP mass value (according to
an increasing mass gap) in order to agree with WMAP data. We
stress that Fig. \ref{DMestimate} does not take into account
several MSSM parameters such as the sfermion masses,
 neutralino composition etc., so it is just a rough estimate that, however, clarifies the role played by the coannihilations.

In the following we will now give a better estimation of
$\langle \s_{eff} v \rangle$ in the two cases $N=2$ and $N=3$
using what we have learnt in our scenario and the coannihilation
cross section of Fig.~\ref{figuracoann} presented in detail in
Appendix A.
\begin{itemize}
 \item $N=2$ case. Assuming that the relative velocities are all equal $v_{ij}\equiv v$ we get:
\begin{equation}
 \langle \s_{eff}^{(2)} v \rangle =\langle\sigma_{22}v\rangle\frac{\langle\sigma_{11}v\rangle/\langle\sigma_{22}v\rangle
+2\langle\sigma_{12}v\rangle/\langle\sigma_{22}v\rangle Q+Q^2}{(1+Q)^2}
\label{sigma2eff}
\end{equation}
where $Q=n_2^{eq}/n_1^{eq}$. The first term in the numerator can
be neglected because the St\"uckelino annihilation cross section
is suppressed by a factor $(C^{(a)})^4$ with respect to the MSSM
neutralino annihilations (see the previous section) and thus
$\langle\sigma_{11}v\rangle\ll \langle\sigma_{22}v\rangle$. The
second term involves the coannihilation cross section. Let us
consider the case in which the NLSP is a generic MSSM neutralino
(a linear combination of $\l^{(1)},\ \l^{(2)}_3,\ \tilde h_d^0,\
\tilde h_u^0$) with a non-vanishing bino or wino components. As we
saw in the previous section each amplitude is generically
proportional to $C^{(a)}g_i$ with $i=1,2$. Without loss of
generality we consider the diagram which involves the bino
component $\Psi_S \l^{(1)}\rightarrow f \bar{f}$ and a photon
exchange in the intermediate channel, i.e the $\M_\g^2$ amplitude
in (\ref{sezcoan}). We get \be C^2_\g= (C^{(1)} \cos\theta_W)^2= 2
(b_2^{(1)})^2 g_{1}^4 \cos^2\theta_W \label{C} \ee From the
expression of the mixed $U(1)'-U(1)_Y - U(1)_Y$ anomaly
(see~\cite{ourpaper}) and from the (\ref{bsds}) we have the
following relation \be b_2^{(1)}=\frac{3(3Q_Q+Q_L)}{256 \pi^2 b_3}
\ee where $b_3=M_{Z'}/4g_0$. With the assumption $M_{Z'}=1$ TeV as
in~\cite{ourpaper} we finally get \be \frac{C^2_\g}{e^2}\simeq
5.76 \times 10^{-12} (3 g_0 Q_Q + g_0 Q_L)^2 \, {\rm GeV}^{-2} \ee
where $e$ is the electric charge. We get similar expressions for
the other three terms in (\ref{sezcoan}). This result has to be
compared to the typical weak cross section
$\langle\sigma_{22}v\rangle\simeq 10^{-9}\, {\rm GeV}^{-2}$. As
long as the charges and the coupling constant of the extra $U(1)$
satisfy the perturbative requirement \be g_0^2\cdot (3 Q_Q +
Q_L)^2<16 \label{chargebound} \ee the following upper bound is
satisfied:
\begin{equation}
 \frac{\langle\sigma_{12}v\rangle}{\langle\sigma_{22}v\rangle}\lesssim 10^{-6}
\label{binosigma12neg}
\end{equation}
in the case of a pure bino, while
\begin{equation}
 \frac{\langle\sigma_{12}v\rangle}{\langle\sigma_{22}v\rangle}\lesssim 10^{-5}
\label{winosigma12neg}
\end{equation}
in the case of a pure wino. Accordingly to
eqs.~(\ref{oh2vssigmaeff}),~(\ref{sigma2eff}),~(\ref{binosigma12neg})
and~(\ref{winosigma12neg}) the relic density gets rescaled
as~\cite{Feldman:2006wd} \be \( \Omega h^2\)^{(2)} \simeq \[
\frac{1+Q}{Q} \]^2 \( \Omega h^2\)^{(1)} \label{2rescaling} \ee We
performed a random sampling of MSSM models in which the NLSP is a
pure bino or a mixed bino-higgsino (the case of a pure wino falls
back into the $N=3$ case due to the wino-chargino mass degeneracy)
and we computed the relic density in presence of coannihilations
using the DarkSUSY package~\cite{Gondolo:2004sc}. These two
situations are easily realized in some corners of the mSUGRA
parameter space. Thus in our scan we assumed this scenario in
order to fix the pattern of the supersymmetry breaking parameters
at weak scale. We emphasize here that this choice is completely
arbitrary, and it is assumed only for simplicity, since in our
model~\cite{ourpaper} the supersymmetry breaking mechanism is not
specified. In the former case there is no model which satisfies
the WMAP constraints~\cite{wmapdata}: \be 0.0913\le\Omega h^2\le
0.1285 \ee since the annihilation cross section of a pure bino is
too low and the rescaling~(\ref{2rescaling}) is not enough to get
the right relic density. In the latter case the higgsino component
tends to increase the annihilation cross section and thus we find
models which satisfy the WMAP constraints. The results are
summarized in Fig.~\ref{DMbino-higgsino} for $\Delta_2=1\%$ and
$\Delta_2=5\%$.
\begin{figure}[t]
      \begin{center}
      \includegraphics[scale=0.7]{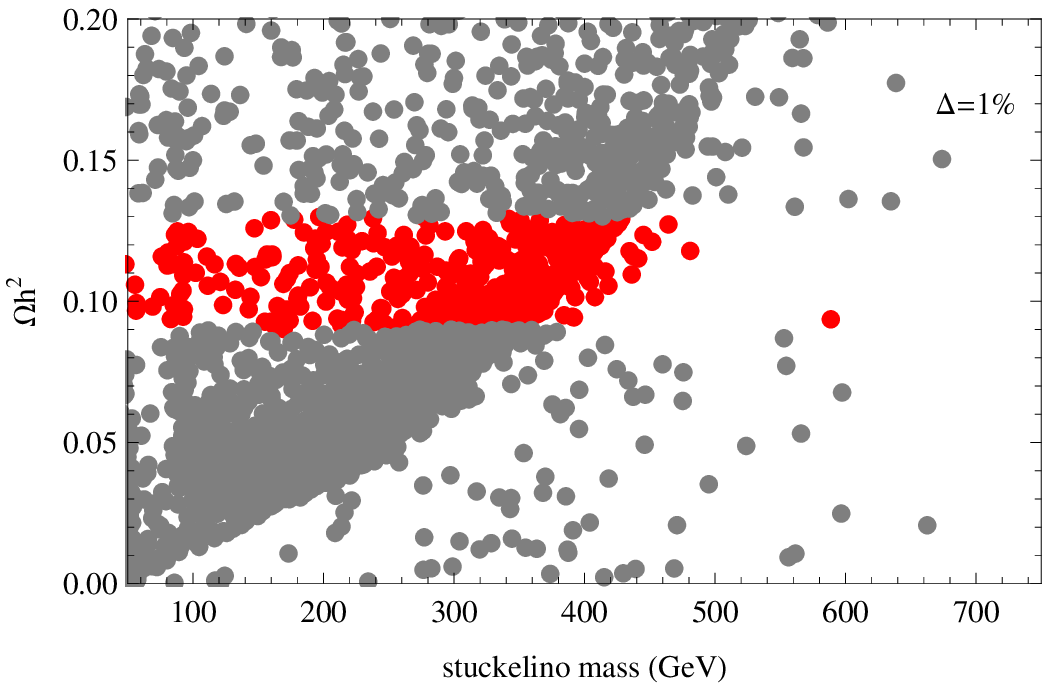}
      \includegraphics[scale=0.7]{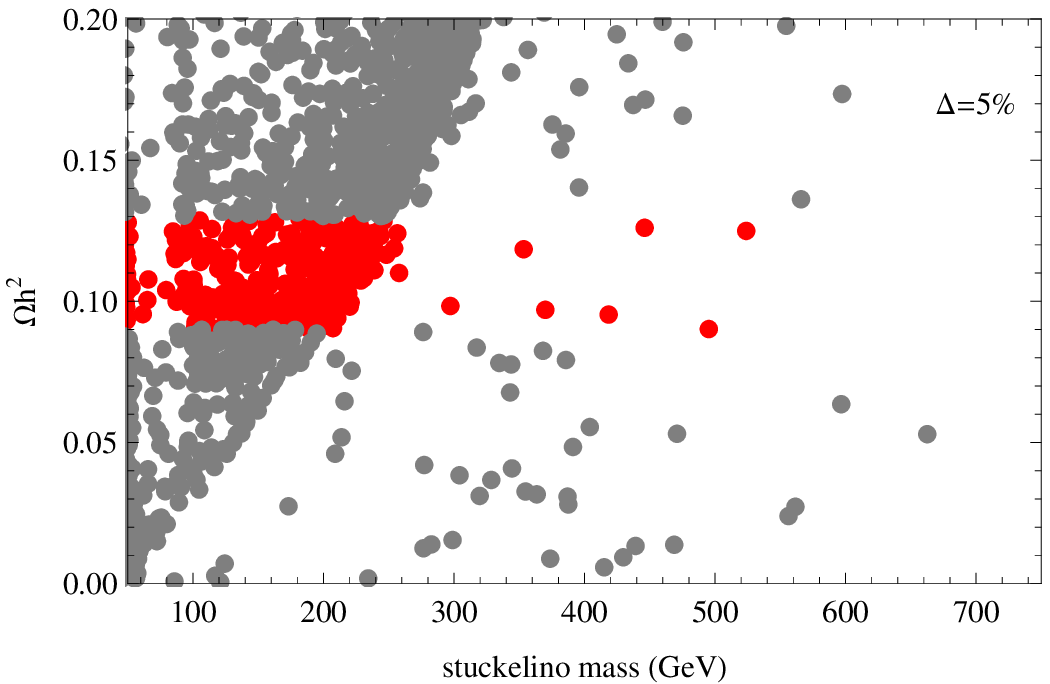}
      \caption{St\"uckelino relic density in the case in which the NLSP is a linear combination bino-higgsino. Red (darker) points denote models which satisfy WMAP data. Left panel: $\Delta_2=1\%$. Right panel:  $\Delta_2=5\%$.}
\label{DMbino-higgsino}
\end{center}
\end{figure}
In order to fulfill the WMAP data (red (darker) points in the
plot~(\ref{DMbino-higgsino})) the St\"uckelino mass must be in the
range $50\; \text{GeV}\lesssim M_S\lesssim 700\; \text{GeV}$ in
the limit $\Delta_2\to 0$, where the lowest bound is given by the
current experimental constraints~\cite{Amsler:2008zzb}.
\item $N=3$ case. This is the case in which there is a third MSSM particle almost degenerate in mass with the LSP and the NLSP.
Typical situations of this kind arise when the NLSP and the next to next to lightest supersymmetric particle
(NNLSP) are respectively the bino and the stau or the wino and the lightest chargino.
Expanding in an explicit way all the terms in the sum~(\ref{sigmaeff}) we get:
\bea
 \la \s_\text{eff}^{(3)} v \ra &=& \la \s_{11} v \ra \g_1^2 + \la \s_{12} v \ra  \g_1 \g_2 + \la \s_{13} v \ra  \g_1 \g_3 +\nn\\
                 &&\la \s_{21} v \ra  \g_2 \g_1 + \la \s_{22} v \ra  \g_2^2 + \la \s_{23} v \ra  \g_2 \g_3 +\nn\\
                 &&\la \s_{31} v \ra  \g_3 \g_1 + \la \s_{32} v \ra  \g_3 \g_2 + \la \s_{33} v \ra  \g_3^2 \nn\\
               &=& \Big[ \la \s_{11} v \ra  (\numeq_1)^2 + \la \s_{12} v \ra  \numeq_1 \numeq_2 + \la \s_{13} v \ra  \numeq_1 \numeq_3 + \nn\\
                 &&\ \la \s_{21} v \ra  \numeq_2 \numeq_1 + \la \s_{22} v \ra  (\numeq_2)^2 + \la \s_{23} v \ra  \numeq_2 \numeq_3 +\nn\\
                 &&\phantom{\Big[}
                  \la \s_{31} v \ra  \numeq_3 \numeq_1 + \la \s_{32} v \ra  \numeq_3 \numeq_2 + \la \s_{33} v \ra  (\numeq_3)^2 \Big]
                 \frac{1}{(\numeq)^2} \nn\\
               &\simeq& %\frac{2 \( \la \s_{12} v \ra  \numeq_1 \numeq_2 + \la \s_{13} v \ra  \numeq_1 \numeq_3 \)}{(\numeq)^2} +\nn\\
%              &&
\frac{ \big[ \la \s_{22} v \ra  (\numeq_2)^2 + 2 \la \s_{23} v \ra  \numeq_2 \numeq_3 +
                 \la \s_{33} v \ra  (\numeq_3)^2 \big] }{(\numeq)^2}
\eea
where in the last line we have neglected the terms $\la \s_{11} v
\ra $, $\la \s_{12} v \ra $ and $\la \s_{13} v \ra $ since these
are the thermal averaged cross sections which involve the
St\"uckelino. By introducing a new set of variables defined by \be
 Q_i=\frac{\numeq_i}{\numeq_1}= \frac{g_i}{g_1} (1 + \D_i)^{3/2} e^{-x_f \D_i} \qquad \text{for} \ i=2,3
\ee
where $g_i$ are the internal degrees of freedom of the particle species, $x_f=m_1/T$ and $\D_i=(m_i-m_1)/m_1$, we obtain
\bea
 \la \s_\text{eff}^{(3)} v \ra
                 %&=& \frac{\la \s_{22} v \ra  Q_2^2 + 2 \la \s_{23} v \ra  Q_2 Q_3 + \la \s_{33} v \ra  Q_3^2}{\(1+Q_2+Q_3\)^2}\nn\\
                 %&&\times\( 1 + \frac{2 \( \la \s_{12} v \ra  Q_2 + \la \s_{13} v \ra  Q_3 \)} {\la \s_{22} v \ra  Q_2^2 + 2 \la \s_{23} v \ra  Q_2 Q_3 + \la \s_{33} v \ra  Q_3^2}\) \nn\\
               &\simeq& \frac{\la \s_{22} v \ra  Q_2^2 + 2 \la \s_{23} v \ra  Q_2 Q_3 + \la \s_{33} v \ra  Q_3^2}{\(1+Q_2+Q_3\)^2}
\eea
Under the assumption $(m_3-m_2)/m_1 \ll 1 /x_f$, $Q_3/Q_2 \simeq g_3/g_2$ we finally get
\be
 \la \s_\text{eff}^{(3)} v \ra \simeq \frac{Q_2^2}{\[1+ \(1 + \frac{g_3}{g_2}\) Q_2\]^2} \la \s_\text{MSSM} v \ra
\label{sigma3eff}
\ee
where
\be
 \la \s_\text{MSSM} v \ra  = \la \s_{22} v \ra  + 2 \frac{g_3}{g_2} \la \s_{23} v \ra  + \(\frac{g_3}{g_2}\)^2 \la \s_{33} v \ra
\ee
In order to compute the rescaling factor between the relic density of our model and the MSSM relic density we have to express $\s_\text{MSSM}$ in terms of a two coannihilating species effective cross section.
This is given by
\bea
 \la \s_\text{eff}^{(2)} v \ra &=& \frac{ \la \s_{22} v \ra  (\numeq_2)^2 + 2 \la \s_{23} v \ra  \numeq_2 \numeq_3 + \la \s_{33} v \ra  (\numeq_3)^2}{(\numeq)^2} \nn\\
               &=& \frac{ \la \s_{22} v \ra  (\numeq_2)^2 + 2 \la \s_{23} v \ra  \numeq_2 \numeq_3 + \la \s_{33} v \ra  (\numeq_3)^2}{(\numeq_2 + \numeq_3)^2} \nn\\
               &=& \frac{ \la \s_{22} v \ra  (\numeq_2)^2 + 2 \la \s_{23} v \ra  \numeq_2 \numeq_3 + \la \s_{33} v \ra  (\numeq_3)^2}{(\numeq_2)^2 (1 + \numeq_3/\numeq_2)^2}\nn\\
               &=& \frac{ \la \s_{22} v \ra   + 2 \la \s_{23} v \ra  Q_{23} + \la \s_{33} v \ra  Q_{23}^2}{ (1 + Q_{23})^2}
\eea
where
\bea
 Q_{23}&=&\numeq_3/\numeq_2= \frac{g_3}{g_2} \(1 + \frac{m_3-m_2}{m_2}\)^{3/2} e^{-x_f \frac{m_3-m_2}{m_2}}\nn\\
       &\simeq& \frac{g_3}{g_2}
\eea
since $(m_3-m_2)/m_1 \ll 1 /x_f$ and $m_2>m_1$ then $(m_3-m_2)/m_2 \ll 1 /x_f$.
We remind the reader that the values of $\numeq_2$, $\numeq_3$ and $\numeq$ are different with respect to those in the former case since now there are only two species in the thermal bath.
We then find
\bea
 \la \s_\text{eff}^{(2)} v \ra &\simeq& \frac{ \la \s_{22} v \ra  + 2 \frac{g_3}{g_2} \la \s_{23} v \ra  + \(\frac{g_3}{g_2}\)^2 \la \s_{33} v \ra  }{ \(1 + \frac{g_3}{g_2}\)^2} \nn\\
                     &\simeq& \frac{\la \s_\text{MSSM} v \ra  }{\(1 + \frac{g_3}{g_2}\)^2}
\eea
and inserting back this relation into~(\ref{sigma3eff}) we obtain
\be
 \la \s_\text{eff}^{(3)} v \ra \simeq \[ \frac{\(1 + \frac{g_3}{g_2}\) Q_2}{1+ \(1 + \frac{g_3}{g_2}\)  Q_2} \]^2 \la \s_\text{eff}^{(2)} v \ra
\ee
The rescaling factor between the three and two particle species relic density is given by the following relation
\be
 \( \Omega h^2\)^{(3)} \simeq \[ \frac{1+ \(1 + \frac{g_3}{g_2}\)  Q_2}{\(1 + \frac{g_3}{g_2}\) Q_2} \]^2 \( \Omega h^2\)^{(2)}\label{rescrelic2to3}
\ee We performed a random sampling of MSSM models with bino-stau
and wino-chargino coannihilations. The first situation is realized
in some corners of the mSUGRA parameter space\footnote{Or in the
so called Constrained MSSM (CMSSM).} while the second situation is
naturally realized in anomaly mediated supersymmetry breaking
scenarios. For each model we computed the relic density $\( \Omega
h^2\)^{(2)}$ for the two coannihilating species with the DarkSUSY
package~\cite{Gondolo:2004sc}. We finally computed $\( \Omega
h^2\)^{(3)}$ using~(\ref{rescrelic2to3}). The bino-stau models
which satisfy the WMAP constraints have a St\"uckelino mass in the
range $100\; \text{GeV}\lesssim M_S\lesssim 350\; \text{GeV}$ in
the limit $\Delta_2\to 0$. As the mass gap increases the number of
allowed models drastically decreases and eventually vanishes for
$\Delta\simeq 5\%$. In the wino-chargino case, models which
satisfy the WMAP constraints are shown in
Fig.~\ref{DMwino-chargino} for four reference values of
$\Delta_2$. The space of parameters with $\Delta_2\lesssim 5\%$
and a St\"uckelino mass $M_S\gtrsim 700$ GeV is favored while as
the mass gap increases lower St\"uckelino masses become favored,
e.g. $100\; \text{GeV} \lesssim M_S<200$ GeV ($\Delta_2\simeq
20\%$).
\begin{figure}[t]
      \begin{center}
      \includegraphics[scale=0.7]{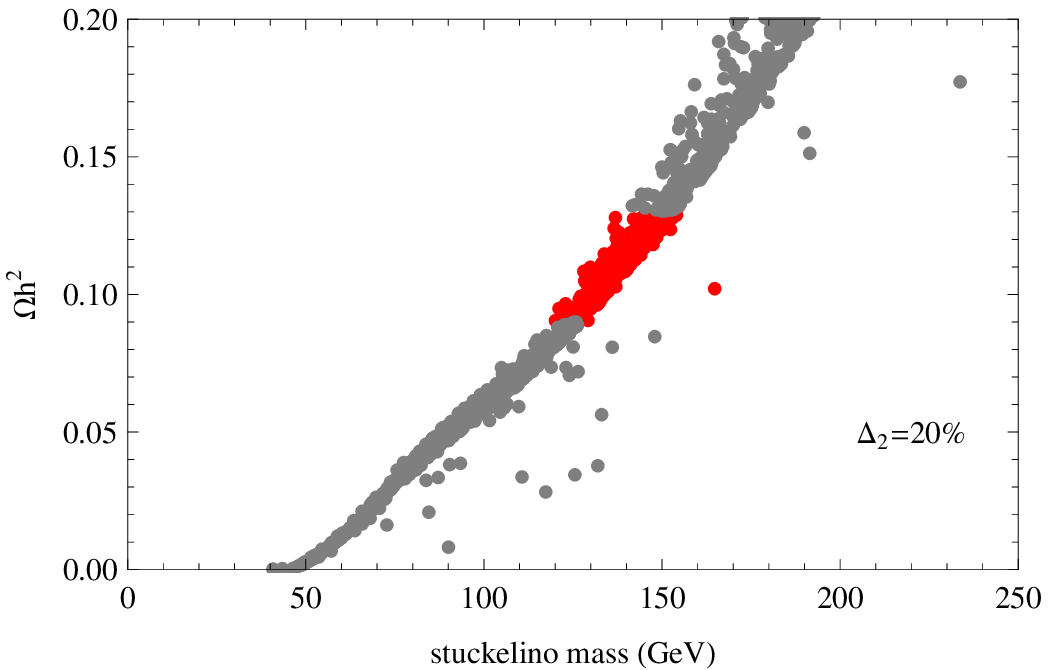}
      \includegraphics[scale=0.7]{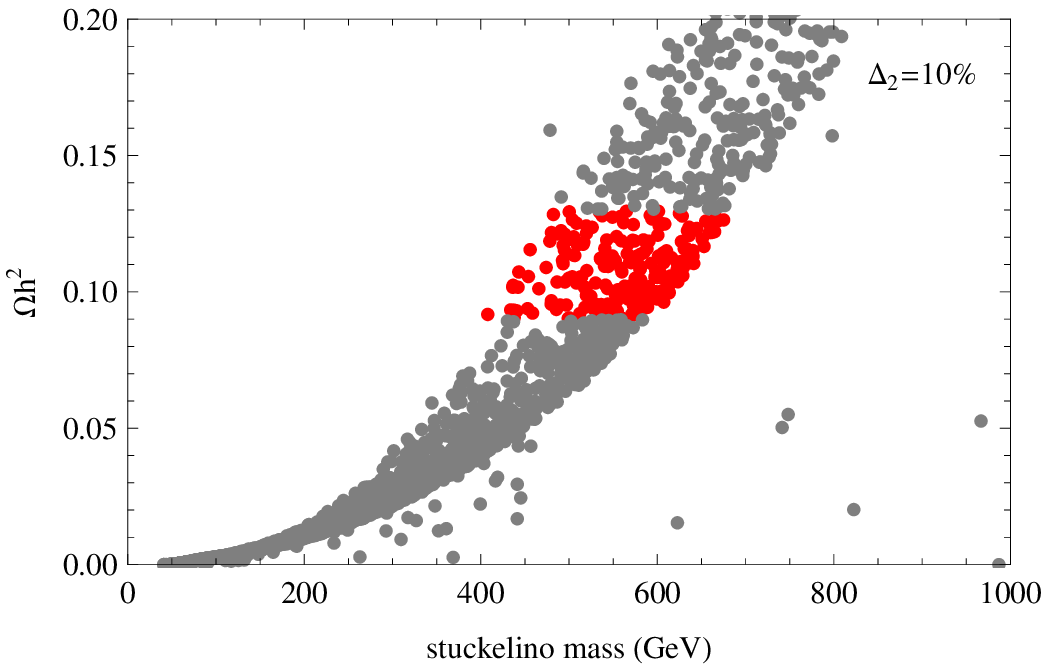}
      \includegraphics[scale=0.7]{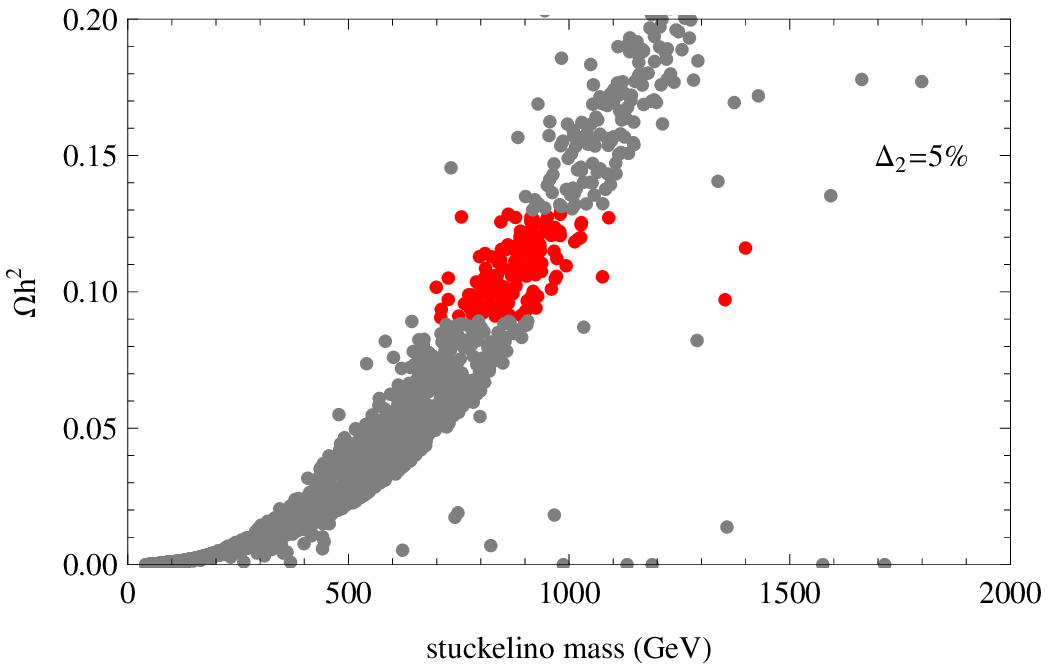}
      \includegraphics[scale=0.7]{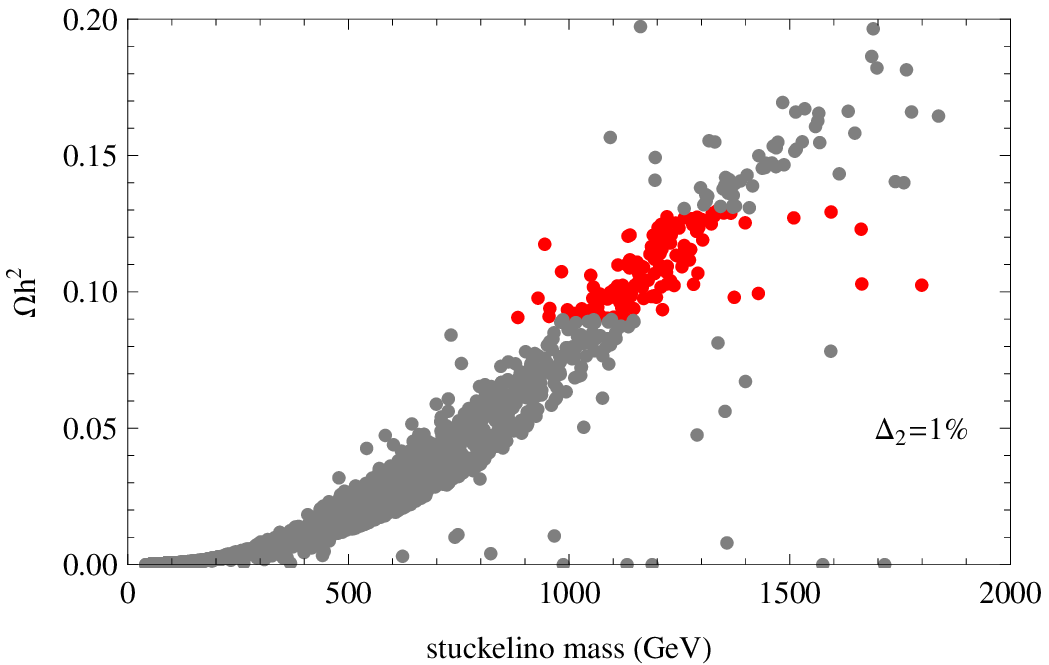}
     \caption{St\"uckelino relic density in the case in which the NLSP is a wino while the NNLSP is the lightest chargino. Red (darker) points denote models which satisfy WMAP data. Upper left panel: $\Delta_2=20\%$. Upper right panel:  $\Delta_2=10\%$. Lower left panel: $\Delta_2=5\%$. Lower right panel: $\Delta_2=1\%$.}
\label{DMwino-chargino}
\end{center}
\end{figure}

\end{itemize}

\section{Conclusions}
We studied a possible dark matter candidate in the framework of
our minimal anomalous $U(1)'$ extension of the
MSSM~\cite{ourpaper}. In the decoupling limit~(\ref{QHuchoice})
and under the assumption $M_0\gg M_S,M_{V^{(0)}}$ the St\"uckelino
turns out to be the LSP. Being an XWIMP the St\"uckelino
annihilation cross section is suppressed with respect to the
typical weak interaction cross sections. This implies that in
order to satisfy the WMAP constraints on the relic density we must
have at least a NLSP almost degenerate in mass with the
St\"uckelino. We considered the case with two and three
coannihilating particles and we found some configuration which
satisfies the WMAP constraints. The results depend on the mass gap
between the St\"uckelino and the NLSP. In the exact degeneracy
limit $\Delta_2\to 0$ the allowed models have a St\"uckelino mass
in the range $50\; \text{GeV}\lesssim M_S\lesssim 700\;
\text{GeV}$ for the bino-higgsino coannihilation case while $900\;
\text{GeV}\lesssim M_S\lesssim 2\; \text{TeV}$ for the
wino-chargino coannihilation case. When the mass gap is
$\Delta_2\simeq 20\%$ the allowed models are those with
wino-chargino coannihilations and a St\"uckelino mass of $100\;
\text{GeV} \lesssim M_S<200$ GeV. Finally let us comment on the
differences between our scenario and that studied in the
work~\cite{Feldman:2006wd}. In our framework the $U(1)'$ does not
arise from a hidden sector and thus all the MSSM fields can be
charged under this extra abelian gauge group. This is the most
relevant feature which could also be detected experimentally (see
for example~\cite{Langacker:2008yv}). Moreover, in our scenario the
St\"uckelino interactions are suppressed with respect to the weak
interactions due to the GS couplings while
in~\cite{Feldman:2006wd} the mechanism to suppress the couplings
and give an XWIMP is provided by the kinetic mixing between the
$U(1)'$ and $U(1)_Y$.

\appendix

\section{Amplitude for $\l_1+\psi_S \to f \bar f$} \label{nnAmp}
In this Appendix we give some details of the amplitude computation for the process $\l_1+\psi_S \to f \bar f$,
\be
 \M = - i k^\m \bar v_S \g_5 \[ \g_\m,\g_\n \] u_1 \[e q_f C_\g \frac{\eta^{\n \r}}{k^2} \bar u_f \g_\r v_f +
                    \frac{g_{Z_0}}{2} C_{Z_0} \frac{\eta^{\n \r}}{k^2-\MZO^2} \bar u_f \g_\r (v_f^{Z_0} - a_f^{Z_0} \g_5) v_f \]
\ee
where $q_f$ denote the electric charges, $v_f^{Z_0}$ and $a_f^{Z_0}$ are the vectorial and axial couplings with $Z_0$,  $C_{\g}=C^{(1)} \cos\theta_W$, $C_{Z_0}=-C^{(1)} \sin\theta_W$ while $k^2=s$ is the momentum of the intermediate gauge boson.
The corresponding square modulus is
\bea
 |\M|^2 &=& - 64 \[
 T_a \(\frac{ a_f C_{Z_0} g_{Z_0} }{k^2-\MZO^2} \)^2
 +T_v \( \frac{ 2 C_\g e q_f}{k^2 }+ \frac{C_{Z_0} g_{Z_0} v_f}{k^2-\MZO^2} \)^2 \]
\label{nunuampl}
\eea
with
\bea
 T_v &=&m_f^4 (p_{\l_1} p_S)
     +M_1 M_S \Big[2 m_f^4+3 (p_f p_{\bar f}) m_f^2+(p_f p_{\bar f})^2\Big]+\nn\\
&&-(p_f p_{\bar f})
   \Big[(p_{\l_1} p_f) (p_f p_S)+(p_{\l_1} p_{\bar f}) (p_{\bar f} p_S)\Big] +m_f^2 \Big[(p_{\l_1} p_S) (p_f p_{\bar f})+\nn\\
&&-2 (p_{\l_1} p_f) (p_f p_S)-(p_{\l_1} p_{\bar f})
   (p_f p_S)-(p_{\l_1} p_f) (p_{\bar f} p_S)-2 (p_{\l_1} p_{\bar f}) (p_{\bar f} p_S)\Big] \nn\\
 T_a &=& \Big[(p_{\l_1} p_{\bar f}) (p_f p_S)+(p_{\l_1} p_f) (p_{\bar f} p_S)\Big] m_f^2-M_1 M_S
   \[m_f^4-(p_f p_{\bar f})^2\]+\nn\\
&&-(p_f p_{\bar f}) \Big[(p_{\l_1} p_f) (p_f p_S)+(p_{\l_1}
p_{\bar f})(p_{\bar f} p_S) \Big] \eea where $p_{\l_1}$, $p_{S}$,
$p_f$ and $p_{\bar f}$ are the bino, St\"uckelino and SM fermions
4-momenta respectively. Writing all the momenta in function of $s$
and integrating over the solid angle we get \bea
&&\!\!\!\!\!\!\!\!\!
\sigma=c_f \(g_1^2 b_2^{(1)}\)^2 \sqrt{s-4 m_f^2} \times \\
&&\!\!\!\!\!\!\!
\times\frac{ \Big[-2 M_1^4+\(4
   M_S^2+s\) M_1^2-6 M_S s M_1-2 M_S^4+s^2+M_S^2 s \Big]}{12 \pi  \(\MZO^2-s\)^2 s^{5/2} \sqrt{M_1^4-2
   \(M_S^2+s\) M_1^2+\(M_S^2-s\)^2}} \times\nn\\
&&\!\!\!\!\!\!\!
\times\Bigg[\(2 m_f^2+s\) \Big(2 \cos\theta_W e q_f \(\MZO^2-s\)
  + \sin\theta_W g_{Z_0} v_f s \Big)^2 +  \(\sin\theta_W g_{Z_0} a_f\)^2 s^2 \(s-4 m_f^2\)\Bigg] \nn
\eea

\begin{comment}
In this Appendix we give the amplitude for the process $\l_1+\psi_S \to \n \n$,
 which is the simplest case over all the possible decays $\l_1+\psi_S \to f \bar f$, since the
neutrinos are massless and are coupled only to the $Z_0$
\be
 \M = -\frac{i}{2} C_{Z_0} g_{Z_0} \frac{k_\m}{k^2 - \MZO^2} \bar v_S
   \g_5 [\g^\m,\g^\n] u_{\l_1} \,
      \bar u_{\n_1} \g_\n (v_\n - a_\n \g_5) v_{\n_2}
\ee
The corresponding square modulus is
\bea
 |\M|^2 &=&
  \sum_\text{spin $\l_1$} \, \sum_\text{spin $\psi_S$} \, \sum_\text{spin
$\n_1$} \,\sum_\text{spin $\n_2$} \M \, \M^* \nn\\
              &=& 64 C_{Z_0}^2 g_{Z_0}^2 \frac{ a_\n^2 + v_\n^2 }{
(k^2-\MZO^2)^2 } (p_{\n_1} \cdot p_{\n_2}) \times\nn\\
             &&\times\Big[ (p_{\l_1} \cdot p_{\n_1}) (p_{\n_1} \cdot
p_{S}) +
                                   (p_{\l_1} \cdot p_{\n_2}) (p_{\n_2}
\cdot p_{S}) -
                                   (p_{\n_1} \cdot p_{\n_2}) M_1 M_S \Big]
\label{nunuampl} \eea where $p_{\l_1}$, $p_{S}$, $p_{\n_1}$ and
$p_{\n_2}$ are the bino, St\"uckelino and neutrinos 4-momenta
respectively and
%we defined $p_\pm = p_{\l_1} \pm p_S$.
$k^2=s$ is the momentum of the intermediate $Z_0$.
The result~(\ref{nunuampl}) is valid only for one family.
If we consider all the families the above amplitude must be multiplied by
$3$.
\end{comment}

%
%\vskip 2cm
%
\begin{flushleft}
{\large \bf Acknowledgements}
\end{flushleft}

\noindent A. R. would like to thank Prof. Michael Green and DAMTP
    for hospitality and the Marie Curie Research Training Network
    ``Superstring Theory'', contract MRTN-CT-2004-512194 and contract ESF JD164 for financial
    support during the completion of this paper. F.Fucito wants to thank N.Fornengo and
P.Ullio for discussions.

\end{document}